\documentclass[a4paper]{jpconf}
\usepackage{graphicx}

\begin{document}
\title{Black holes in modified gravity theories
}

\author{A. de la Cruz-Dombriz,
A. Dobado 
and A. L. Maroto,
}%

\address{Departamento de  F\'{\i}sica Te\'orica I, Universidad Complutense
 de Madrid, 28040 Madrid, Spain.
}
\ead{dombriz@fis.ucm.es}
\begin{abstract}
In the context of $f(R)$ gravity theories, the issue of finding
static and spherically symmetric black hole solutions is addressed. 
Two approaches to study the existence of such solutions are considered: 
first, constant curvature solutions,
and second, the general case (without imposing constant curvature) 
is also studied. 
Performing a perturbative expansion around the Einstein-Hilbert action,
it is found that only solutions of the Schwarzschild-(Anti-) de Sitter type
are present (up to second order in perturbations) and the explicit expressions for 
these solutions are provided 
in terms of the $f(R)$ function.
Finally we consider the thermodynamics of black holes in Anti-de Sitter
space-time and study their local and global stability.
%
\end{abstract}

\section{Introduction}
In the last years, increasing attention has been paid to modified theories of gravity
in order to understand several open cosmological issues such as the present accelerated
expansion of the universe \cite{Carroll}. 
Some of those theories modify general relativity by adding higher powers of the scalar curvature $R$, i.e. $f(R)$ theories, the
Riemann and Ricci tensors or their derivatives \cite{Maroto&Dobado_1993}.
It is therefore quite natural to ask about black hole (BH) features in those gravitational
theories since, on the one hand, some BH signatures may be peculiar to Einstein's gravity and others may be robust features of all generally covariant theories of gravity. On the other hand,
the obtained results may lead to rule out some models which will be in desagreement with
expected physical results. For thoses purposes, research on thermodynamical quantities
of BH is of particular interest.

In this work we will restrict ourselves to the so called $f(R)$ gravity
theories
in metric formalism in Jordan's frame where
the gravitational Lagrangian is given by  $R+f(R)$
and modified Einstein's equations are
usually fourth order in the metric.
%
%
Using the euclidean action method \cite{Hawking&Page, Witten}
in order to determine different thermodynamical quantities,
Anti-de Sitter ($AdS$) BHs in $f(R)$ models
and their stability will be studied \cite{Dombriz_BH}.
\section{Black Holes in $f(R)$ gravities}
The most general static and spherically symmetric $D\geq 4$
dimensional external metric for the gravitational field produced
by a nonrotating object in $f(R)$ gravity theories can be written
as:
\begin{eqnarray}
d s^2\,=\,e^{-2\Phi(r)} A(r) d t^2-A^{-1}(r) d r^2-r^2 d\Omega_{D-2}^2 \,\,\,\equiv\,\,\,\lambda(r) d t^2-\mu^{-1}(r) d r^{2}-r^2 d\Omega_{D-2}^2
\label{metric_D_v1}
\end{eqnarray}
where $d\Omega_{D-2}^2$ is the metric on the $S^{D-2}$ sphere.
%
%
%
%
In this case the scalar curvature $R$ in $D$ dimensions depends only on  $r$ and
for the first parametrization in equation (\ref{metric_D_v1}) is given by
\begin{eqnarray}
R=\frac{1}{r^2}\left\{(D-2)[(D-3)(1-A)+2r(-A'+A\Phi')]+r^{2}[3A'\Phi'-A''-2A(\Phi'^2-\Phi'')]
\right\}.
\label{Dcurv}
\end{eqnarray}
where the prime denotes derivative with respect to $r$ from now on.
Therefore the most general static and spherically symmetric 
metrics with constant scalar curvature $R_{0}$ can be found 
solving the equation $R=R_0$. For a constant $\Phi(r)=\Phi_{0}$
it is straightforward to see that the solution is:
\begin{eqnarray}
A(r)\,=\,1+a_{1}r^{3-D}+a_{2}r^{2-D}-\frac{R_0}{D(D-1)}r^2
\label{A_solution_R_constant_Dobado_procedure}
\end{eqnarray}
with $a_{1,2}$ being arbitrary integration constants. For
the particular case $D=4$, $R_{0}=0$ and $\Phi_{0}=0$, the
metric can be written exclusively in terms of the function
$A(r)\,=\,1+a_{1}r^{-1}+a_{2}r^{-2}$ where
by establishing the identifications $a_{1}=-2G_{N}M$ and
$a_{2}=Q^2$, this solution corresponds to a charged massive BH solution with mass $M$
and charge $Q$ (Reissner-Nordstr\"{o}m solution).

On the other hand,  if the constant curvature case $R=R_0$
is considered, the modified Einstein's equation from $f(R)$ theories for the function $A(r)$
reduces to:
%
\begin{equation}
R+f(R)+(1+f_R)\left[A''+\frac{(D-2)}{r}\left(A'-2A\Phi'\right)
-3A'\Phi'+2A\left(\Phi'^2-\Phi''\right)\right]\,=\,0
\label{eq_motion_A}
\end{equation}
where $f_R$ holds for $df(R)/dR$ .
As it is explained in \cite{Dombriz_BH}, the vacuum constant curvature solutions of
$f(R)$ gravities are given by:
\begin{equation}
R_0=\frac{D\,f(R_0)}{2(1+f_{R_0})-D}
\label{const}
\end{equation}
If the hypothesis
%
$\Phi'(r)=0$ 
is made, then equation (\ref{eq_motion_A}) , by using (\ref{const}), reduces in this case to
\begin{equation}
A''+(D-2)\frac{A'}{r}=-\frac{2}{D}R_0 \,\,\,\rightarrow\,\,\,  A(r)\,=\,c_1\,+\,c_2r^{3-D}-\frac{R_0}{D(D-1)}r^2
\label{A_eq_R_constant}
\end{equation}
where
$c_{1,2}$ are two arbitrary constants. However
this solution has no constant curvature in
 the general case since, as it can be seen 
in expression (\ref{A_solution_R_constant_Dobado_procedure}),
the constant curvature requirement
demands $c_{1}=1$. Then, for negative $R_0$, this solution
is basically the $D$ dimensional generalization obtained by Witten \cite{Witten}
of the BH in $AdS$ space-time solution considered by Hawking and Page 
\cite{Hawking&Page}. With the natural choice $\Phi_0=0$ the
solution can be written as:
\begin{eqnarray}
A(r)\,=\,1-(R_{S}/r)^{D-3}+r^2/l^2
\label{A_solution}
\end{eqnarray}
where $R_{S}$ and $l$ have the usual interpretations found in the literature.

Thus we have concluded  that the only static and
spherically symmetric vacuum solutions with constant (negative) curvature of any
$f(R)$ gravity is just the Hawking-Page BH in $AdS$ space.
However this kind of solution is not the most general static and
spherically symmetric metric with constant curvature as can be seen by
comparison with the solutions given by expression
(\ref{A_solution_R_constant_Dobado_procedure}).
%
\section{Perturbative results}
In Einstein-Hilbert (EH) theory the most general static
solution with spherical symmetry is the one with constant curvature. 
However, this is not guaranteed to be the case
in $f(R)$ theories. The problem of finding the general static spherically
symmetric solution in arbitrary $f(R)$ theories without imposing the
constant curvature condition is in principle too complicated. For that reason
a perturbative analysis of the problem is performed here, assuming that
the modified action is a small perturbation around EH theory of the form
$f(R)\,=\,-(D-2)\Lambda_{D}+\alpha\,g(R)$
where $\alpha\ll 1$ is a dimensionless parameter and $g(R)$ is assumed
to be analytic.
%
%
Assuming that the $\lambda(r)$ and
$\mu(r)$ functions appearing in the second parametrization of the metric (\ref{metric_D_v1}) 
are also analytical in $\alpha$, they can be written as follows
\begin{eqnarray}
\lambda(r)\,=\,\lambda_{0}(r)+\sum_{i=1}^{\infty}\alpha^{i}\lambda_{i}(r)  \,\,\,\,;\,\,\,\, 
\mu(r)\,=\,\mu_{0}(r)+\sum_{i=1}^{\infty}\alpha^{i}\mu_{i}(r)
\label{expansion_en_alpha_lambda&mu}
\end{eqnarray}
where $\{\lambda_{0}(r),\,\mu_{0}(r)\}$ are the unperturbed solutions for the
EH action with cosmological constant $\Lambda_D$ given by
$\mu_{0}(r)\,=\,1+\frac{C_1}{r^{D-3}}-\frac{\Lambda_{D}}{(D-1)}\,r^2$
and $\lambda_{0}(r)\,=\,-C_{2}(D-2)(D-1)\,\mu_{0}(r)$
which are the standard BH solutions in a $D$ dimensional $AdS$
spacetime with constant curvature $R_0\equiv D\Lambda_{D}$. 

By inserting the proposed $f(R)$ expansion
 and equation (\ref{expansion_en_alpha_lambda&mu}) in modified Einstein's equations
the first and second order equations in $\alpha$ are obtained, whose solutions are
\begin{eqnarray}
&&\lambda_{1}(r)=(D-1)(D-2)\left\{
C_{4}+\frac{C_{1}C_{4}-C_{2}C_{3}}{r^{D-3}}
-\left[\frac{C_{4}\Lambda_{D}}{D-1}+\frac{C_{2}A(g;\,D,\,\Lambda_D)}{(D-1)(D-2)}\right] r^{2}\right\}\nonumber\\
%
&&\mu_{1}(r)=\frac{C_{3}}{r^{D-3}}+\frac{A(g;\,D,\,\Lambda_{D})}
{(D-2)(D-1)}r^{2}\nonumber\\
&&\lambda_{2}(r)=C_{6}+\frac{C_{5}+C_{6}C_{1}}{r^{D-3}}
-\left[\frac{C_{6}\Lambda_{D}}{D-1}-A(g;\,D\,\Lambda_{D})
\left(C_{4}+C_{2} g_{R_0}-\frac{2 C_{2} R_{0}g_{R_{0}R_{0}}}{D-2}\right)\right] r^2
\nonumber\\
&&\mu_{2}(r)=\left(C_{3}C_{4}+\frac{-C_{5}}{(D-2)(D-1)}\right)\frac{C_{2}^{-1}}{r^{D-3}}+\frac{A(g;\,D,\,\Lambda_{D})
\left(2 R_{0} g_{R_{0}R_{0}}-(D -2) g_{R_0}\right)}{(D -2)^2 (D -1)}r^{2}
\end{eqnarray}
where $g_{R_0}\equiv dg(R)/dR|_{R=R_0}$, $g_{R_{0}R_{0}}\equiv d^{2}g(R)/dR^{2}|_{R=R_0}$, 
and
$A(g;\,D,\,\Lambda_{D})\equiv g(R_0)-2\Lambda_D g_{R_0}$.
Higher order solutions $\{\lambda_{3,4,...}(r),\mu_{3,4,...}(r) \}$ 
can be obtained by inserting
the previous results in the corresponding equations order $\alpha^{3,4,...}$ 
which become increasingly complicated.

Notice that from the obtained results up to second order in $\alpha$,
the corresponding metric has constant scalar curvature for any
value of the parameters $C_{1,\dots,6}$. As a matter of fact,
this metric is nothing but the standard Schwarzschild-$AdS$ geometry,
and can be easily rewritten in the usual form by making a
trivial time reparametrization.
%
Therefore, at least up to second order, the only static,
spherically symmetric solutions which are analytical
in $\alpha$ are the standard Schwarzschild-$AdS$ space-times.
\section{Black-hole thermodynamics}
For simplicity
we will concentrate only on constant curvature $AdS$ BH solutions given by expression (\ref{A_solution})
with $\Phi(r)\equiv 0$ as a natural choice.
Then, usual definitions of temperature lead to
$\beta\equiv 1/T=4 \pi l^2 r_{H}/[(D-1)r_H^2+(D-3)l^2]$
where $r_H$ satisfies $A(r_H)=0$.
Notice that the temperature is a function of $r_H$ only, i.e. it
depends only on the BH size.
It is easy to prove that $T$ has a
minimum $T_0$ at $r_{H0}$ and whose
%
existence was established in \cite{Hawking&Page}
for $D=4$.
Notice that for any $T > T_0$, there are two
possible BH sizes: one corresponding to the small BH phase with
$r_H < r_{H0}$ and the other corresponding to the large BH phase with $r_H > r_{H0}$.
The remaining thermodynamic quantities
are determined by using the Euclidean method
so that the free energy $F$ is given by
\begin{equation}
F=-\frac{(R_0+f(R_0)) \mu_{D-2}}{36 \pi (D-1)
G_D}(l^2r^{D-3}_H-r^{D-1}_H)
\end{equation}
where 
provided $-(R_0+f(R_0))>0$, which is the usual
case in EH gravity, $F>0$ for $r_H<l$ and $F<0$ for $r_H>l$. 
%
%
On the other hand, 
%
the heat capacity $C$ can be written as:
\begin{equation}
C\,=\,\frac{\partial E}{\partial T}=\frac{\partial E}{\partial
r_H}\frac{\partial r_H}{\partial T}\,=\,\frac{-(R_0+f(R_0))(D-2)\mu_{D-2}r^{D-2}_Hl^2}{8G_D(D-1)}
\frac{(D-1)r^2_H+(D-3)l^2}{(D-1)r^2_H-(D-3)l^2}.
\end{equation}

Let us assume from now on that, like in the EH case,  the condition
$(R_0+f(R_0))<0$ holds. This is the required condition, as explicitly shown in \cite{Dombriz_BH} to get both positive energy and entropy  in these theories and according to (\ref{const}) it implies $1+f_{R_0}>0$ in agreement with cosmological requirements \cite{Silvestri}. With this assumption,
%
%
we find $C>0$ for $r_H > r_{H0}$ (the large BH region) and $C<0$ for
$r_H < r_{H0}$ (the small BH region). For $r_H \sim r_{H0}$ ($T$
close to $T_0$)   $C$ is divergent. 
In any case, for $f(R)$ theories with $R_0+f(R_0)<0$, we have found
a similar scenario to the one described in full detail
in \cite{Hawking&Page} for the EH case:
%
%
for $T>T_0$ two possible BH
solutions are present, the small BH and the large BH: The
small one has $C<0$ and $F>0$ as the
standard Schwarzschild BH.
For the large BH two possibilities appear; if
$T_0<T<T(r_H=l)$ then $C>0$ and $F>0$ and the BH will decay by tunneling into radiation, but if
$T>T(r_H=l)$ then $C>0$ and $F<0$.
In this case $F$ of the large BH will
be less than that of pure radiation. 
%
\\
\\
To illustrate the previous calculations, a model
$f(R)\,=\,\alpha (-R)^{\beta}$ is considered where
the heat capacity $C$ and the free energy $F$ signs will be studied to know
the local and global stability of BHs. This model has
a nonvanishing constant curvature solution
$R_{0}\,=\,-\left[(2-D)/\alpha(2\beta-D)\right]^{1/(\beta-1)}$
calculated by using (\ref{const}) and therefore
two separated regions, where $R_{0}<0$ holds, have to be studied:
Region $1$ $\{\alpha<0,\,\beta>D/2\}$ and Region $2$ $\{\alpha>0,\, \beta<D/2\}$.
Further models following an analogous procedure were presented in 
\cite{Dombriz_BH}. 
\begin{center}
\begin{figure}[h]
\begin{minipage}{12pc}
\includegraphics[width=12pc]{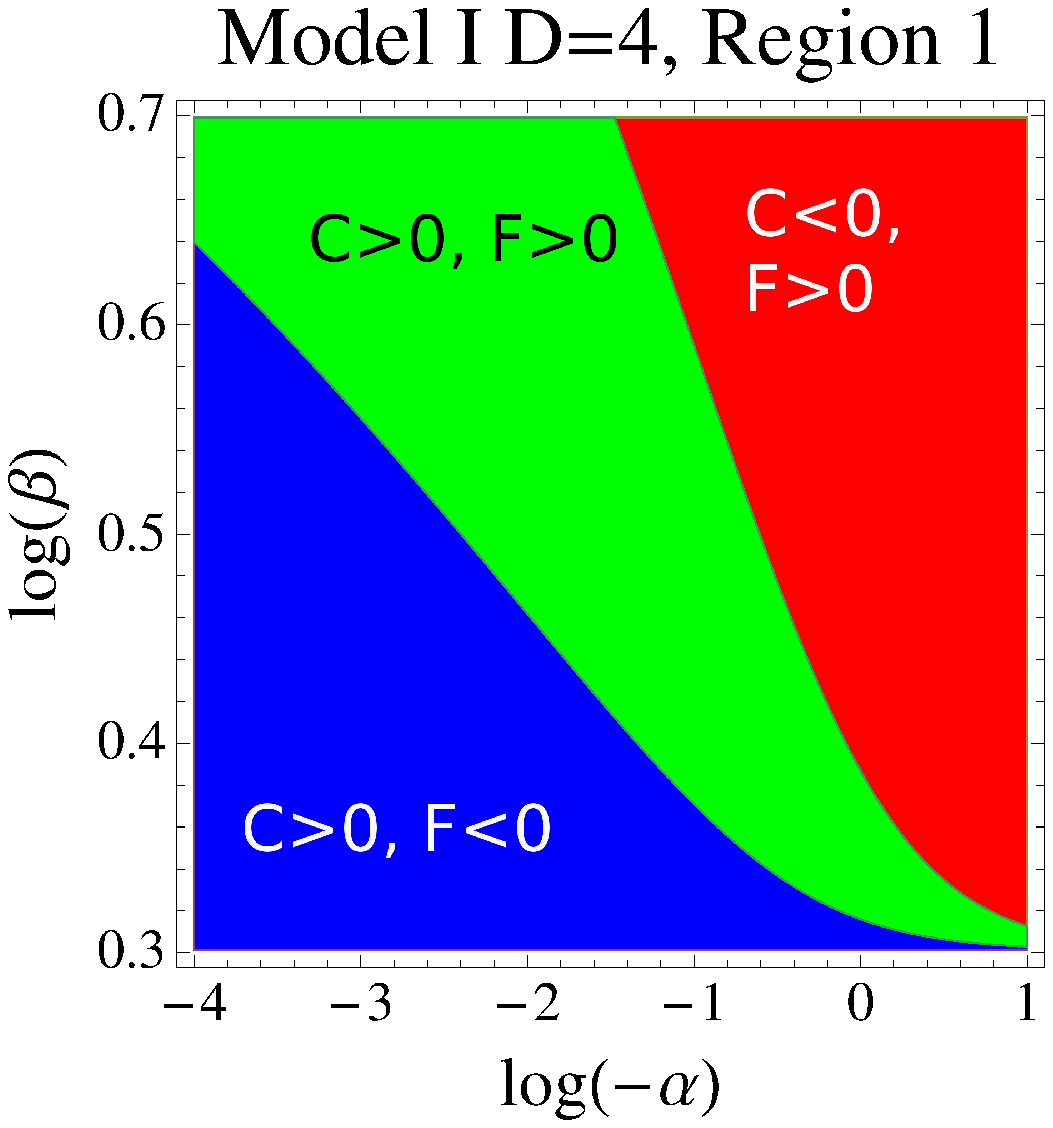}
\end{minipage}\hspace{1.5pc}%
\begin{minipage}{12pc}
\includegraphics[width=13.2pc]{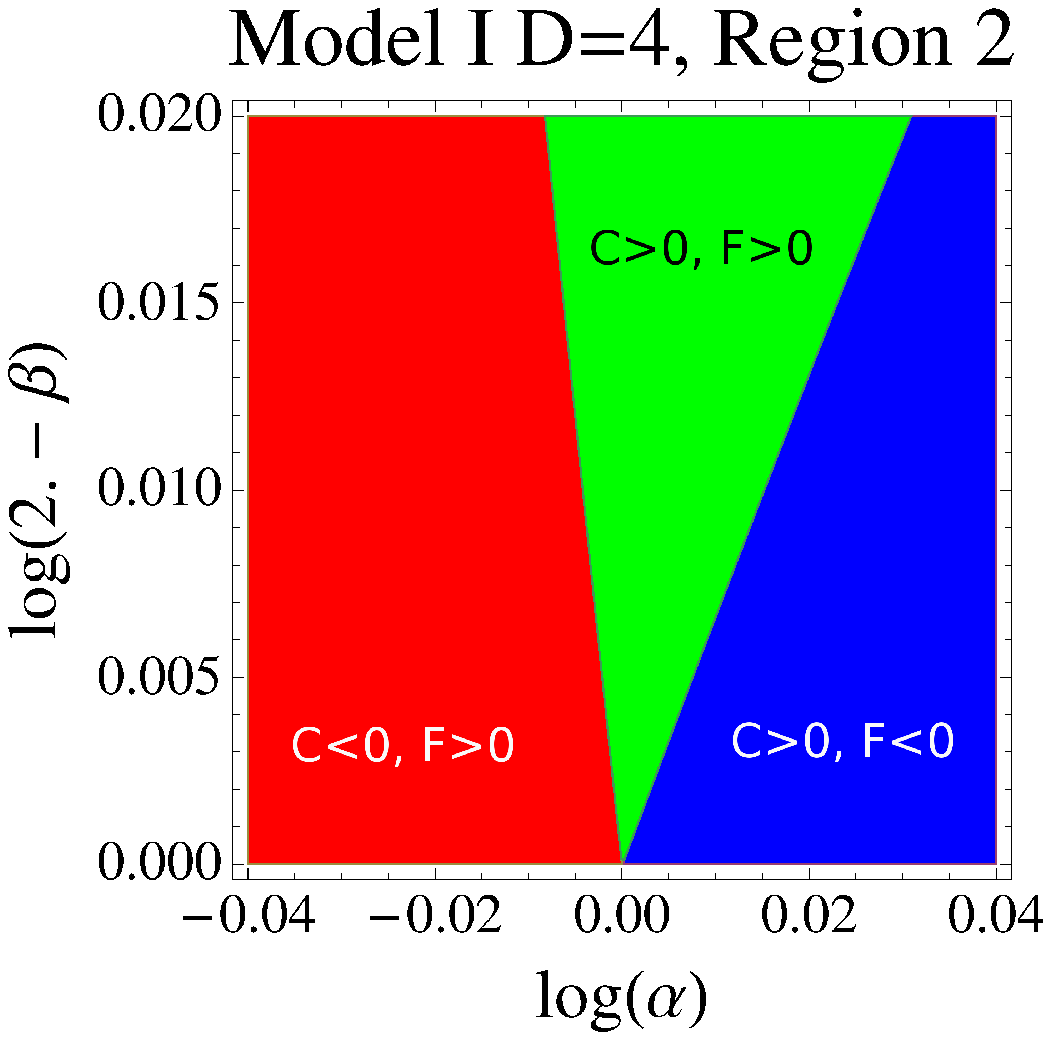}
\end{minipage} 
\caption{Regions for $C$ and $F$ signs
have been compared in the $(\alpha,\beta)$ plane for Model I in $D=4$.
Region 1(left), Region 2 (right). Schwarzschild radius 
$R_S=2$ was assumed.}
\end{figure}
\end{center}
\ack This work has been supported by Ministerio de Ciencia e Innovaci\'on (Spain) project numbers FIS 2008-01323 and FPA 2008-00592, UCM-Santander PR34/07-15875 and UCM-BSCH GR58/08 910309.
\\


\end{document}